\documentclass[a4paper,11pt]{article}
\pdfoutput=1 

\usepackage{jcappub} 

\usepackage{graphicx}
\usepackage{bm}
\usepackage{amsmath}
\usepackage{amssymb}
\usepackage{xcolor}
\usepackage{hyperref}
\usepackage[capitalise]{cleveref}
\usepackage[activate={true,nocompatibility},final,kerning=true,factor=1100,stretch=10,shrink=10]{microtype}
\usepackage{academicons}
\usepackage{xcolor}
\usepackage{fontawesome5}

\definecolor{orcidlogocol}{rgb}{0.65, 0.807, 0.223}

\newcommand{\orcid}[1]{$\,$\href{https://orcid.org/#1}{\textcolor{orcidlogocol}{\faOrcid}}}

\usepackage[T1]{fontenc} 

\title{\boldmath Primordial Black Hole stellar microlensing constraints: understanding their dependence on the density and velocity distributions}

  \author{Anne M. Green \orcid{0000-0002-7135-1671}}
  \affiliation{School of Physics and Astronomy, \\ University of Nottingham
 \\Nottingham NG7 2RD, United Kingdom}

\emailAdd{anne.green@nottingham.ac.uk} 

\abstract{Microlensing surveys of stars in the Large Magellanic Cloud constrain the fraction of the Milky Way halo in Primordial Black Holes (PBHs) with mass  $10^{-9} \lesssim M/M_{\odot} \lesssim 10^{4}$. Various studies have reached different conclusions on the uncertainties in these constraints due to uncertainties in the Dark Matter (DM) distribution. We therefore revisit the dependence of the microlensing differential event rate, and hence exclusion limits, on the DM density and velocity distributions. The constraints on the abundance of low- and high-mass PBHs depend, respectively, on the long- and short-duration tails of the differential event rate distribution. Long-duration events are due to PBHs moving close to the line of sight and their rate (and hence the constraints on low-mass PBHs) has a fairly weak ($\sim 10\%$) dependence on the DM density and velocity distributions. Short-duration events are due to PBHs close to the observer and their rate (and hence the constraints on moderate- and high-mass PBHs) depends much more strongly on the DM velocity distribution. An accurate calculation of the local DM velocity distribution is therefore crucial for accurately calculating PBH stellar microlensing constraints.}

\begin{document}
\maketitle
\flushbottom

\section{\label{sec:intro}Introduction}

Primordial Black Holes (PBHs), black holes formed from the collapse of large overdensities in the early Universe~\cite{ZN,Hawking:1971ei}, are a potential dark matter (DM) candidate, and the detection of gravitational waves from mergers of black holes by LIGO-Virgo~\cite{LIGOScientific:2016aoc} has invigorated interest in PBH DM~\cite{Bird:2016dcv,Clesse:2016vqa,Sasaki:2016jop,Carr:2016drx,Carr:2020xqk,Green:2020jor}.  Stellar microlensing, the temporary brightening of a star when a compact object (CO), such as a PBH, crosses the line of sight~\cite{Paczynski:1985jf,Griest:1990vu}, is one of the most powerful tools for probing planetary or Earth mass COs.

In the 1990s the  EROS~\cite{Palanque-Delabrouille:1997cxg}, MACHO~\cite{MACHO:1995udp} and OGLE~\cite{Wyrzykowski:2009ep} collaborations began monitoring millions of stars in the Magellanic Clouds in order to probe COs in the Milky Way (MW) halo. The typical duration of a Large Magellanic Cloud (LMC) microlensing event caused by a CO with mass $M$ is $ \sim (M/M_{\odot})^{1/2}  \, {\rm yr}$, see Eq.~(\ref{tref}). Therefore a long-exposure survey is required to probe multi-solar mass COs, and a high-cadence survey to probe planetary mass COs. The EROS-2 survey~\cite{EROS-2:2006ryy} excluded COs with mass $10^{-7} \lesssim M/M_{\odot} \lesssim 10$ making up all of the DM in the MW halo. In recent years the upper end of the excluded mass range has been extended to $\sim 10^{3} M_{\odot}$ by combining data from EROS-2 and MACHO~\cite{Blaineau:2022nhy} and then to $\sim 10^{4} M_{\odot}$ by the OGLE-III and -IV long-exposure surveys~\cite{Mroz:2024mse,Mroz:2024wag}. Very recently the minimum excluded mass has been reduced to $10^{-9} M_{\odot}$ by the OGLE high-cadence survey~\cite{Mroz:2024wia}.

Constraints almost always involve modelling and assumptions. In the case of LMC stellar microlensing, the differential event rate, and hence the constraints on the halo fraction, $f$, of COs depend on the CO density and velocity distribution in the MW~\cite{Griest:1990vu}. Various studies (e.g. Refs.~\cite{MACHO:1995udp,Hawkins:2015uja,Green:2017qoa,Calcino:2018mwh,Garcia-Bellido:2024yaz}) have found that the uncertainty in the constraints is significant. The tightest limit on $f$ varies by a factor of a few, while the largest value of $M$ for which $f=1$ is excluded can vary by more than an order of magnitude. 
Some authors argue that the uncertainties are sufficiently large that multi-Solar mass PBHs making up all of the DM is in fact not excluded~\cite{Carr:2023tpt,Garcia-Bellido:2024yaz}.
On the other hand, the OGLE collaboration finds that the uncertainty in the constraint from their long-exposure survey is relatively small (see Extended Data Fig.~1 of Ref.~\cite{Mroz:2024wag}).

Motivated by this apparent discrepancy, we revisit the dependence of the microlensing differential event, and constraints on the halo fraction in COs, on the DM distribution.  Our aim is to understand how the constraints depend on the DM density and velocity distributions, in order to elucidate what is required for a reliable calculation of the constraints. In Sec.~\ref{sec:background} we present the halo models that we use, their differential event rates and the exclusion limits on the halo fraction of COs from a toy long-exposure microlensing survey.  In Sec.~\ref{sec:physical} we explore the origin of the behaviour of the microlensing differential event rate, 
and  the accuracy of assuming a Maxwellian velocity distribution. Finally, we conclude with summary and discussion in Sec.~\ref{sec:summary}

\section{Models}
\label{sec:background}

In Sec.~\ref{subsec:eventrate} we present the calculation of the microlensing differential event rate for the standard halo model and the power law halo models we consider, before calculating the exclusion limits for these models from a toy long-exposure microlensing survey in Sec.~\ref{subsec:constraints}.

\subsection{Differential event rate}
\label{subsec:eventrate}

The duration of a microlensing event, $\hat{t}$, can be defined as the time taken to cross the Einstein {\it diameter}~\footnote{This is the definition used by the MACHO collaboration. The EROS and OGLE collaborations instead define the duration as the Einstein radius crossing time.}, $\hat{t} = 2 R_{\rm E}/v_{\rm t}$, where $v_{\rm t}$ is the transverse speed of the lens relative to the line of sight and $R_{\rm E}(x)$ is the Einstein radius
\begin{equation}
R_{{\rm E}}(x) = 2 \left[ \frac{ G M x (1-x)L}{c^2 } \right]^{1/2} \,, \label{re}
\end{equation}
where $G$ is the Gravitational constant, $M$ is the mass of the CO, $L \approx 50 \, {\rm kpc}$ is the distance to the LMC and $x$ is the distance of the lens from the observer in units of $L$ ($0 \leq x \leq 1$). It will be useful later to define a reference timescale 
\begin{equation}
\label{tref}
t_{\rm ref} \equiv \frac{4}{v_{\rm c}} \left( \frac{G M L}{c^2} \right)^{1/2} \sim 1 \, {\rm yr} \left( \frac{M}{1 M_{\odot}} \right)^{1/2} \,,
\end{equation}
where $v_{\rm c}  \approx 220 \, {\rm km \, s}^{-1}$ is the circular speed,
so that 
\begin{equation}
\hat{t} = \frac{v_{\rm c}}{v_{\rm t}} \sqrt{ x (1-x)}   t_{\rm ref}  \,.
\end{equation}

The standard expression for the LMC microlensing differential event rate, ${\rm d} \Gamma/ {\rm d} \hat{t}$, for a halo composed entirely of COs with mass $M$ and a smooth density distribution $\rho(r)$ is~\cite{MACHO:1996qam}:  
\begin{equation}
\label{df}
\frac{{\rm d} \Gamma}{{\rm d} \hat{t}} =  \frac{32 L}
                 { M {\hat{t}}^4
              v_{{\rm c}}^2}
              \int^{1}_{0} \rho(x) R^{4}_{{\rm E}}(x)
              \exp{\left[-Q(x)\right]}  {\rm d} x \,, 
\end{equation}
where  $Q(x)= 4 R^{2}_{{\rm E}}(x) / (\hat{t}^{2} v_{{\rm c}}^2)$. Equation (\ref{df}) assumes that the velocity distribution, $f({\bf v})$, is Maxwellian 
\begin{equation}
\label{maxwellian}
f({\bf v}) \, {\rm d}^3 {\bf v} = \frac{1}{\pi^{3/2} v_{\rm c}^{3}} \exp{ \left( -\frac{v^2}{v_{\rm c}^2} \right)} \, {\rm d}^{3} {\bf v} \,, 
\end{equation}
however this assumption is only exact if the density varies with Galactocentric radius $r$ as $\rho(r) \propto r^{-2}$ for all $r$. Since the MW halo is in reality finite the velocity distribution should be truncated at the (Galactocentric radius dependent) escape speed, $v_{\rm esc}(r)$, however (compared to other uncertainties) this has a relatively small effect on exclusion limits (see below). 

We assume that the MW DM distribution is smooth. PBHs formed from the collapse of large density perturbations will be more clustered on sub-galactic scales than particle dark matter, due to the Poisson fluctuations in their initial distribution~\cite{Afshordi:2003zb}. If the perturbations are gaussian then the clusters are sufficiently extended~\cite{Inman:2019wvr,Jedamzik:2020ypm} that the PBHs still act as lenses individually and the effect on the microlensing event rate is negligible (apart, possibly, for the largest CO masses probed by long-exposure surveys)~\cite{Petac:2022rio,Gorton:2022fyb}. Non-gaussianity can lead to enhanced clustering~\cite{Young:2019gfc}, however, and if the clusters are so compact that the cluster as a whole acts as a lens then the microlensing constraints would be shifted to smaller PBH masses~\cite{Calcino:2018mwh}.

The standard halo (SH) model usually assumed in microlensing studies (`Model S' in e.g. Ref.~\cite{MACHO:1996qam}) is a cored isothermal sphere with a density profile
\begin{equation}
\rho(r) = \rho_{0} \, \frac{r_{{\rm c}}^2 + r_{0}^2}{r_{{\rm c}}^2 + r^2} \,,
\label{rhor}
\end{equation}
and local dark matter density $\rho_{0} \equiv \rho(r_{0}) = 0.0079 M_{\odot} {\rm pc}^{-3}$, core radius $r_{{\rm c}} = 5$ kpc, 
Solar radius $r_{0} = 8.5$ kpc and circular speed $v_{\rm c}  =220 \, {\rm km \, s}^{-1}$. Some of the best-fit values of the parameters appearing in this model have changed in recent years, e.g.~the Solar radius has been measured as $r_{\rm 0} = (8.18 \pm 0.01 \pm 0.02) \, {\rm kpc}$ by the GRAVITY collaboration~\cite{2019A&A...625L..10G}. 
These changes have a relatively small effect on the microlensing differential event rate compared with changes in the density profile~\cite{Green:2017qoa,Calcino:2018mwh}. A reliable calculation of the exclusion limits on COs is our eventual goal. However, the focus of this paper is understanding the dependence of the differential event rate and exclusion limits on the DM distribution. Therefore, to facilitate comparison with past work, we use the standard values of the parameters.
The differential event rate, Eq.~(\ref{df}), for the SH model is given by~\cite{MACHO:1996qam} 
\begin{equation}
\label{dfsh}
\frac{{\rm d} \Gamma}{{\rm d} \hat{t}} = \frac{512 \rho_{0} 
          (R_{{\rm c}}^2 + R_{0}^2) L G^2  M }
             {  {\hat{t}}^4 {v_{{\rm c}}}^2 c^4}
              \int^{1}_{0} \frac{x^2 (1-x)^2}{A + B x + x^2}
            \exp{\left[ -Q(x)\right] }{\rm d} x \,, 
\end{equation}
where $A=(r^2_{{\rm c}}+ r^2_{0})/L^2$, $B=-2(r_{0}/L) \cos{b}
\cos{l}$ and $b = -32.8^{\circ}$ and $l = 281^{\circ}$ are the Galactic
latitude and longitude, respectively, of the LMC.  

{Evans'  power law (PL) models~\cite{1994MNRAS.267..333E} have tractable expressions for the differential event rate, which include the (non-Maxwellian) velocity distribution (see Appendix B of Ref.~\cite{MACHO:1994ovs}). They are therefore often used as benchmarks for studying the dependence of the microlensing differential rate on the DM density distribution (e.g.~Refs.~\cite{Evans:1994rv, MACHO:1994ovs,MACHO:1995udp,Green:1999vh,Hawkins:2015uja,Green:2017qoa}). These models are self-consistent, in that the velocity distribution is calculated from the potential of the halo. However close to the MW disc (i.e.~at small line of sight distances $x$), the disc potential is non-negligible and will affect the velocity distribution, in particular increasing the velocity dispersion.  The PL halo models have a density profile~\footnote{We focus on the spherical case, $q=0$, in Eq.~(2) of Ref.~\cite{1994MNRAS.267..333E}.}
\begin{equation}
\label{rhoPL}
\rho(r) = \frac{v_{\rm a}^2 r_{\rm c}^{\beta}}{4 \pi G} \frac{ (1-\beta^2) r^2 +  3 r_{\rm c}^2 }{ (r^2 + r_{\rm c}^2)^{(\beta + 4)/2} } \,,
\end{equation}
where $v_{\rm a}$ is a normalization velocity and $\beta$ is constant. For $r \ll r_{\rm c}$, $\rho(r) \rightarrow$ const while for $r \gg r_{\rm c}$, $\rho(r) \propto r^{-(2+ \beta)}$ i.e.~for $\beta > (<) \, 0$ the density decreases more (less) rapidly with increasing $r$ than for an isothermal sphere. We consider models B and C from Refs.~\cite{MACHO:1995udp}. Model B has a total mass within 50 kpc $M(r<50 \, {\rm kpc})=7 \times 10^{11} \, M_{\odot}$, with a rising rotation curve ($\beta = -0.2$) with normalization velocity $v_{\rm a} = 200 \, {\rm km \, s}^{-1}$, while Model C has a light halo, $M(r<50 \, {\rm kpc})=2 \times 10^{11} \, M_{\odot}$, with a falling rotation curve ($\beta = 0.2$) and $v_{\rm a}= 180 \, {\rm km \, s}^{-1}$. Both models are spherical and have a core radius $r_{\rm c} = 5 \, {\rm kpc}$. The normalization velocity determines the overall depth of the potential well, and hence the typical velocities of the COs, and the numerical values were chosen to give a total rotation speed at $r_{0}$ within $15\%$ of $220\, {\rm km \, s}^{-1}$~\cite{MACHO:1994ovs}.  Figure \ref{fig:vc} shows the rotation curve of the SH and PL halos B and C. 

\begin{figure}
    \centering
    \includegraphics[width=0.8\textwidth]{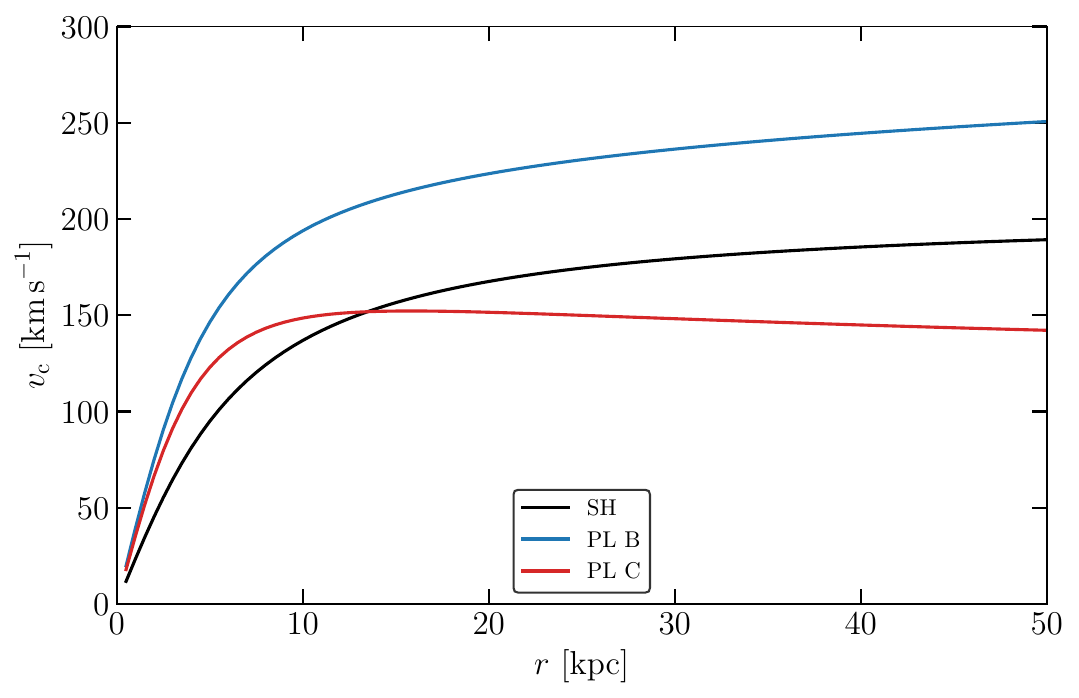}
    \caption{The rotation speed, $v_{\rm c}$, as a function of radius, $r$, for the standard halo (black) and power law halos B (blue) and C (red).
    \label{fig:vc}}
\end{figure}

\begin{figure}
    \centering
    \includegraphics[width=0.8\textwidth]{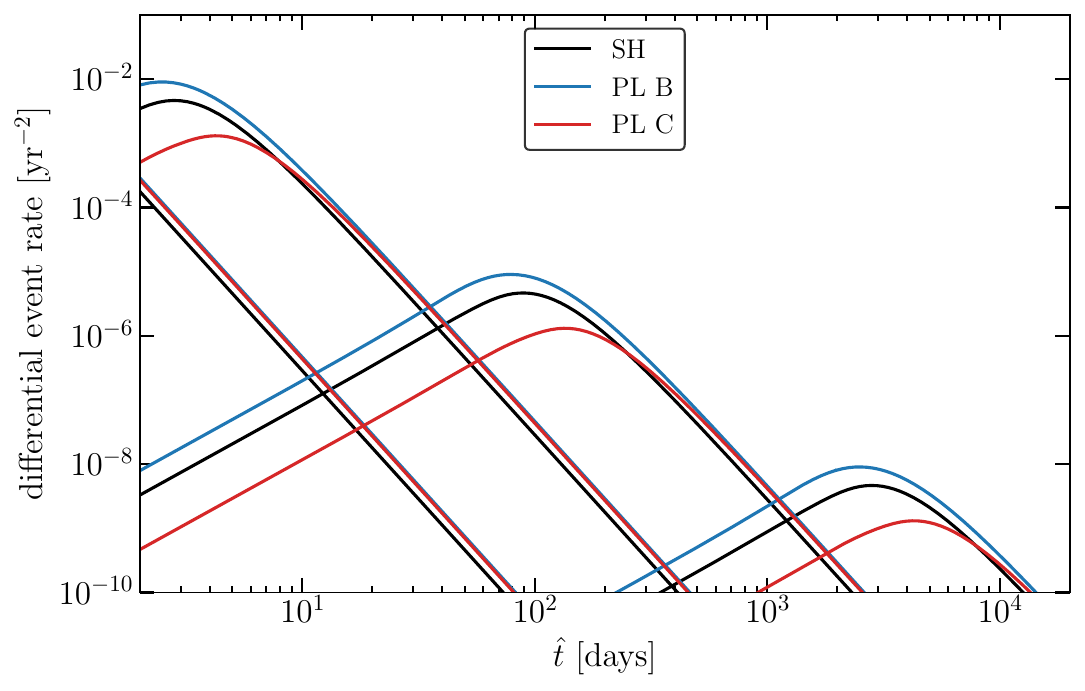}
    \caption{The differential microlensing event rate towards the  LMC, ${\rm d} \Gamma/ {\rm d} \hat{t}$, as a function of Einstein diameter crossing time, $\hat{t}$, if the Milky Way halo is composed entirely of compact objects with mass (from left to right) $M= 10^{-6}, 10^{-3}, 1$ and $10^{3} M_{\odot}$.
The standard halo model is shown in black and power law halos B and C in blue and red respectively (see Sec.~\ref{subsec:eventrate}  for details of the models).
    \label{fig:ratevarym}}
\end{figure}

Figure \ref{fig:ratevarym} shows the differential microlensing event rate towards the  LMC, ${\rm d} \Gamma/ {\rm d} \hat{t}$, as a function of Einstein diameter crossing time, $\hat{t}$, for the SH and PLs B and C. We assume all of the halo is in COs of a single mass, $M$,  and show the differential event rate for four masses which span the range that a long-exposure survey sensitive to events with durations $ 1 \,  {\rm day} \lesssim \hat{t} \lesssim 10^{4} \, {\rm day}$ can probe: $M= 10^{-6}, 10^{-3}, 1$ and $10^{3} M_{\odot}$. As pointed out by Griest~\cite{Griest:1990vu}, the differential event rate scales as $M^{-1/2}$ and the durations as $M^{1/2}$. In this plot, however, we show a range of masses, to make explicit how a given survey is sensitive to different regions of the differential event rate distribution for different masses. A long-exposure survey, will probe the long-duration tail ($\hat{t} \gg t_{\rm ref}$)  of the distribution for light compact objects ($M \lesssim 10^{-4} M_{\odot}$) and the short-duration tail ($\hat{t} \ll t_{\rm ref}$) for heavy compact objects ($M \gtrsim 10^{3} M_{\odot}$).
As found in Refs.~\cite{MACHO:1995udp,Zhao:1995qi}, for short-duration events ${\rm d} \Gamma / {\rm d} \hat{t} \propto \hat{t}^{2}$ while for long-duration events ${\rm d} \Gamma / {\rm d} \hat{t} \propto \hat{t}^{-4}$, independent of the DM distribution. For these 3 models, the amplitude of the short-duration tail varies by more than an order of magnitude, with PL B (C) being larger (smaller) than the SH by a factor of $\sim 2.4 \,  (7.1)$. The maximum differential event rate for PL B (C) is a factor of 1.9 (3.6) larger (smaller) than for the SH and the event duration at which it occurs is a factor of 1.2 (1.5) smaller (larger) for PL B (C) than for the SH. In the case of the long-duration tail, the PL halos have a similar amplitude, and the SH is smaller by a factor of $1.6$. We will explore the physical origin of the differences in the amplitudes of the short- and long-duration tails in Sec.~\ref{sec:physical} below. 

The origin of the power law scalings of the rate of long- and short-duration events was explained by Mao and Paczysnski~\cite{Mao:1996hr}. The duration of an event is $\hat{t} = 2 R_{\rm E} /v_{\rm t}$, where $v_{\rm t}$ is the tangential velocity which is given by $v_{\rm t} = v \sin{i}$, where $v$ is the relative velocity between the lens and the observer and $i$ is the angle of inclination between the velocity and the line of sight. Long-duration events are caused by lenses which move almost along the light of sight. In this case $\hat{t} \propto \sin^{-1}{i} \propto {i}^{-1}$. The total number, $N$, (rather than differential rate) of such events is proportional to the solid angle, $(1- \cos{i})$, divided by $\hat{t}$, so that $N \propto i^2/i^{-1} \propto i^3 \propto \hat{t}^{-3}$. Short-duration events are caused by lenses that are very close to either the observer ($x \sim 0$) or lens ($x \sim 1$) where the Einstein radius, $R_{\rm E}(x)$ is small.  For lenses close to the observer $\hat{t} \propto R_{\rm E} \propto [x L]^{1/2}$ and the total number of these events is $N \propto R_{\rm E} x L/ \hat{t}  \propto \hat{t}^{3}$, with a similar argument leading to the same scaling for lenses close to the source.

\subsection{Constraints}
\label{subsec:constraints}

\begin{figure}
    \centering
    \includegraphics[width=0.8\textwidth]{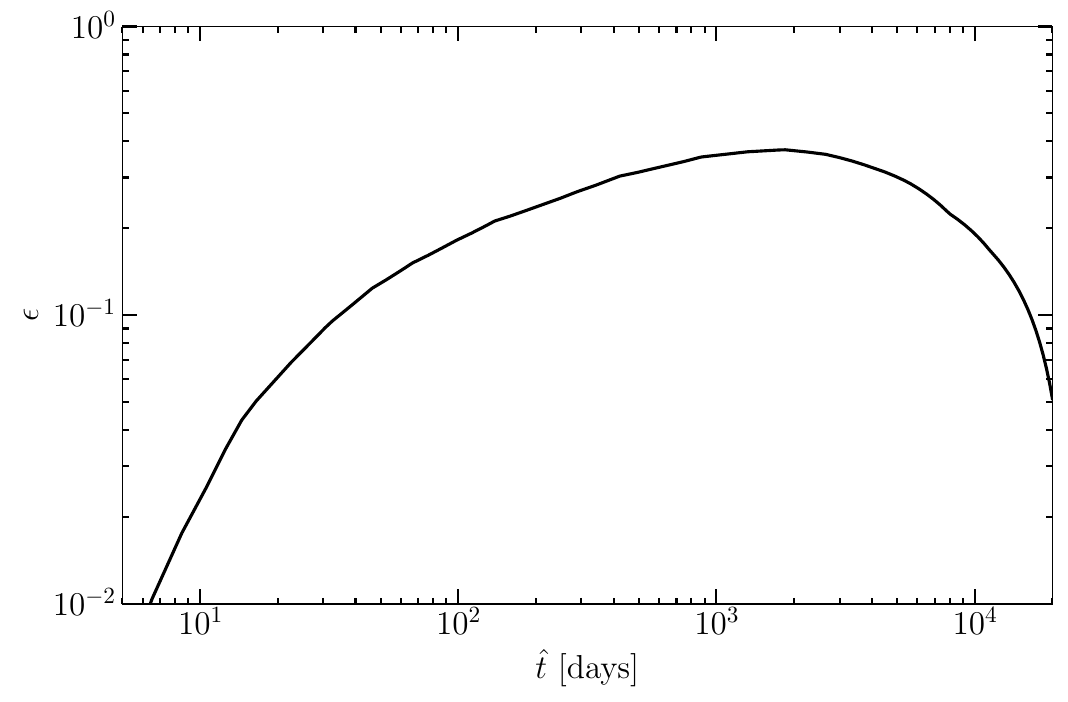}
    \caption{The efficiency, $\epsilon$, as a function of Einstein diameter crossing time, $\hat{t}$, of our toy long-exposure microlensing survey.  
    \label{fig:eff}}
\end{figure}

The expected number of events, $N_{\rm exp}$, is given by
\begin{equation}
\label{nexp}
N_{{\rm exp}} = E \int_{0}^{\infty} \frac{{\rm d} \Gamma}{{\rm d} \hat{t}}
           \,  \epsilon(\hat{t}) \, {\rm d} \hat{t} \,,         
\end{equation}
where $E$ is the exposure in star years and $\epsilon(\hat{t})$ is the detection efficiency i.e.~the probability that a microlensing event that occurs with duration $\hat{t}$ is detected. For concreteness, we consider a toy long-exposure microlensing survey, which roughly mimics the OGLE 20-year survey~\cite{Mroz:2024wag}~\footnote{Our conclusions would also apply, qualitatively, to a short-exposure, high-cadence microlensing survey that is sensitive to lighter CO, such as Ref.~\cite{Mroz:2024wia}.}. We assume that 79 million stars are observed for 7000 days, giving an exposure of $E=5.5 \times 10^{11}$ star days. We use the efficiency for observing field number 501 from Ref.~\cite{OGLEdata}, which is shown, as a function of Einstein diameter crossing time, in Fig.~\ref{fig:eff}.
We calculate exclusion limits as for the `strict' case in Ref.~\cite{Mroz:2024mse}, assuming that none of the observed events are due to dark COs~\footnote{If instead some of the observed events are allowed to be due to COs, the contribution of stars to the differential event rate has to be included, and the exclusion limit on $f$ is weakened by a factor of a few for masses $10^{-4} \lesssim M/M_{\odot} \lesssim 10$.}. For smoothly distributed COs the probability distribution of the number of observed events is Poissonian. Therefore, if no events are observed, the $95\%$ confidence limit on the fraction of the halo in COs is given by $f=3.0/N_{{\rm exp}}$. 

Figure \ref{fig:flimshPL} shows the resulting constraint on the halo fraction, $f$, as a function of CO mass, $M$, for the standard halo model and power law halos B and C (the three models for which the differential event rate is shown in Fig.~\ref{fig:ratevarym}). The constraints are tightest for $10^{-3} \lesssim M/M_{\odot} \lesssim 1$, where the differential event rate is large for durations for which the efficiency is largest ($10^{2} \, {\rm day} \lesssim \hat{t} \lesssim 10^{4} \, {\rm day}$). For $M \gtrsim M_{\odot}$ the toy long-exposure survey is sensitive to the short-duration tail of the differential event rate and the order of magnitude variation in its amplitude leads to an order of magnitude variation in the constraint on $f$ for a given $M$. The largest mass for which $f=1$ is excluded is a factor of $1.9$ ($3.5$) larger (smaller) for PL B (C) than for the SH.  For $M \lesssim 10^{-4}M_{\odot}$ the toy long-exposure survey is sensitive to the long-duration tail of the differential event rate, and hence the variation in the constraint on $f$ is smaller than for larger $M$. The smallest mass for which $f=1$ is excluded is a factor of $\sim 1.6$ smaller for both PL B and C than for the SH.  The tightest limit on $f$ is a factor of $1.7$ ($1.5$) lower (higher) for PL B (C) than for the SH. These results qualitatively match those previously found in Refs.~\cite{MACHO:1995udp} and \cite{Green:2017qoa} for the MACHO first-year and EROS-2 surveys respectively. 
We have checked that truncating the velocity distribution at a constant $v_{\rm esc}$ has a relatively small effect on the constraints on $f$
for the SH.  The constraints are tighter at small $M$ and weaker at larger $M$, by of order $1\%$ for $v_{\rm esc} = 500 \, {\rm km \, s}^{-1}$  and of order $10\%$ for $v_{\rm esc} = 400 \, {\rm km \, s}^{-1}$ (the MW escape at the Solar radius is likely in the range $(500-550) \, {\rm km \, s}^{-1}$~\cite{2021A&A...649A.136K}).

\begin{figure}
    \centering
    \includegraphics[width=0.8\textwidth]{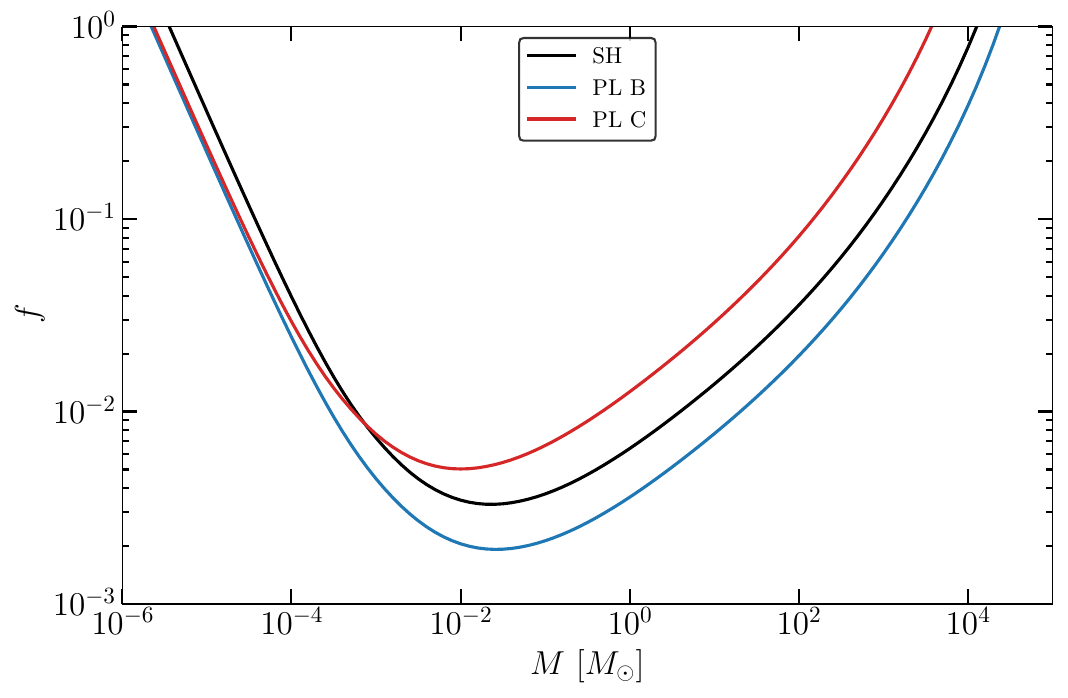}
    \caption{The constraint on the halo fraction, $f$, of compact objects as a function of CO mass, $M$, from a toy long-duration LMC microlensing survey for the standard halo model (black) and power law halos B (blue) and C (red). 
    \label{fig:flimshPL}}
\end{figure}

\section{Behaviour of differential event rate}
\label{sec:physical}

In Sec.~\ref{sec:tails} we explore the physical origin of the differences in the amplitude of the long- and short-duration tails of the differential event rate observed in Sec.~\ref{subsec:eventrate}, while in Sec.~\ref{sec:gaussian} we explore the accuracy of assuming a Maxwellian velocity distribution for the PL halos.

\subsection{Tails of differential event rate}
\label{sec:tails}

It is useful to rewrite the expression for the differential event rate for a smooth DM distribution with density profile, $\rho(r)$, and a Maxwellian velocity distribution, Eq.~(\ref{df}), as 
\begin{equation}
\label{eventraterewrite}
\frac{{\rm d} \Gamma}{{\rm d} \hat{t}} = \frac{512 L^3 G^2  M \rho_{0}}
             {  {\hat{t}}^4 {v_{{\rm c}}}^2 c^4}
              \int^{1}_{0} \tilde{\rho}(x) x^2 (1-x)^2 
            \exp{\left[ - Q(x) \right]}     {\rm d} x \,, 
\end{equation}
with
\begin{equation}
Q(x) =  x(1-x) \left( \frac{t_{\rm ref}}{\hat{t}} \right)^2 \,,
\end{equation} 
where $t_{\rm ref}$ is given by Eq.~(\ref{tref}) and the density is written in terms of the local density, $\rho_{0}$, as $\rho(x)= \rho_{0} \tilde{\rho}(x)$, so that $\tilde{\rho}(x)$ is dimensionless and encodes the variation of the density along the line-of-sight, and (by definition) $\tilde{\rho}(0)=1$.

\subsubsection{Short-duration tail}
\label{sec:short}
We saw in Sec.~\ref{subsec:eventrate} that the amplitude of the $\hat{t}^2$ short-duration ($\hat{t} \ll t_{\rm ref}$) tail varies significantly. As explained by Mao and Paczysnski~\cite{Mao:1996hr}, short-duration events are caused by lenses which are close to either the observer ($x \approx 0$) or source ($x \approx 1$). For $\hat{t} \ll t_{\rm ref}$,  unless $x \approx 0$ or $x \approx 1$, $Q(x)$ is large and  $\exp{[-Q(x)]} \approx 0$. Defining $T=(t_{\rm ref} / \hat{t} )^2 \gg 1$ and 
\begin{equation}
\left( \frac{{\rm d} \Gamma}{{\rm d} \hat{t}} \right)_{0} = \frac{512 L^3 G^2  M \rho_{0}}
             {  {\hat{t}}^4 {v_{{\rm c}}}^2 c^4} \,,
             \end{equation}
the expression for the differential event rate, Eq.~(\ref{eventraterewrite}), can therefore be approximated by 
\begin{eqnarray} 
\frac{{\rm d} \Gamma}{{\rm d} \hat{t}} & = & \left( \frac{{\rm d} \Gamma}{{\rm d} \hat{t}} \right)_{0} \left\{  \int_{0}^{1/2}   \tilde{\rho}(x) x^2 (1-x)^2 
            \exp{\left[ - x(1-x) T \right]}   \,  {\rm d} x  \right.  \nonumber \\
            &&  +   \left. \int_{1/2}^{1}  \tilde{\rho}(x) x^2 (1-x)^2 
            \exp{\left[ -x (1-x) T \right]}   \,   {\rm d} x        \right\} \,, \nonumber \\
            & \approx &  \left( \frac{{\rm d} \Gamma}{{\rm d} \hat{t}} \right)_{0}  \left[  \int_{0}^{1/2}   x^2  
            \exp{\left( - x T \right)}   \,  {\rm d} x  +  \tilde{\rho}(1)  \int_{1/2}^{1}   (1-x)^2 
            \exp{\left( -(1-x) T \right)}   \,  {\rm d} x        \right] \,, \nonumber \\
               & \approx &  \left( \frac{{\rm d} \Gamma}{{\rm d} \hat{t}} \right)_{0} \left[   \int_{0}^{1/2}   x^2  
            \exp{\left( - x T \right)} \,    {\rm d} x  +  \tilde{\rho}(1)  \int_{0}^{1/2}   y^2 
            \exp{\left( -y T \right)}  \,   {\rm d} y        \right] \,, \nonumber \\
            & \approx & \left( \frac{{\rm d} \Gamma}{{\rm d} \hat{t}} \right)_{0} \left[   \int_{0}^{\infty}   x^2  
            \exp{\left( - x T \right)}  \,   {\rm d} x  +  \tilde{\rho}(1)  \int_{0}^{\infty}   y^2 
            \exp{\left( -y T \right)}   \,  {\rm d} y        \right] \,,  \nonumber \\
          &   = & \left( \frac{{\rm d} \Gamma}{{\rm d} \hat{t}} \right)_{0} \left[  1  +   \tilde{\rho}(1) \right] \frac{2}{T^3} \approx  \left( \frac{{\rm d} \Gamma}{{\rm d} \hat{t}} \right)_{0} \frac{2}{T^3} = \frac{\rho_{0} v_{\rm c}^4 c^2 \hat{t}^2}{4 M^2 G}  \,, 
\label{shorteventrateapprox} 
\end{eqnarray}
where in the final line we use the fact that the DM density at the LMC is much less than the local DM density, $\tilde{\rho}(1) \ll 1$.
For the SH model Eq.~(\ref{shorteventrateapprox}) agrees with the full expression, Eq.~(\ref{eventraterewrite}), at the $5\%$ level (and at the $1\%$ level if the contribution from lenses close to the source, $\tilde{\rho}(1)$, is included).  The amplitude of the short-duration tail is determined by the product of the local dark matter density and the fourth power of the circular speed.  A larger circular speed means the typical speed of the COs is larger. This increases the rate at which the COs enter the microlensing tube and also decreases the duration of the events they cause.

\subsubsection{Long-duration tail}
We saw in Sec.~\ref{subsec:eventrate} that the amplitude of the $\hat{t}^{-4}$ long-duration ($\hat{t} \gg t_{\rm ref}$) tail is slightly larger for PLs B and C than for the SH. As explained by Mao and Paczysnski~\cite{Mao:1996hr}, long-duration events are caused by lenses which move almost along the light of sight. For $\hat{t} \gg t_{\rm ref}$, $Q(x) \approx 0$ and $\exp{[-Q(x)]} \approx 1$ so that the expression for the differential event rate assuming a Maxwellian velocity distribution, Eq.~(\ref{eventraterewrite}), becomes
\begin{equation}
\label{dflongduration}
\frac{{\rm d} \Gamma}{{\rm d} \hat{t}} = \frac{512 L G^2  M \rho_{0}}
             {  {\hat{t}}^4 {v_{{\rm c}}}^2 c^4}
              \int^{1}_{0} \tilde{\rho} (x) x^2 (1-x)^2  {\rm d} x \,, 
\end{equation}
with the integral being independent of $\hat{t}$. The total number of COs within the microlensing tube is determined by the local density, $\rho_{0}$, and the integral of $\tilde{\rho} (x)$ within the tube. The fraction of COs moving along the line of sight depends on the normalization of the speed distribution, which for the Maxwellian, Eq.~(\ref{maxwellian}), is inversely proportional to $1/v_{\rm c}^2$.  Consequently, the amplitude of the long-duration tail is proportional to $\rho_{0} / v_{\rm c}^2$. The circular speed is given by $v_{\rm c}^2 = G M(<r)/ r$, where $M(<r)$ is the mass enclosed within a radius $r$. If only the DM is considered $v_{\rm c}^2 \propto \rho_{0}$ and the amplitude of the long-duration tail is independent of $v_{\rm c}$ and $\rho_{0}$ (physically a larger DM density leads to a wider spread in velocities, and hence less COs with $v_{\rm t} \approx 0$). Consequently, the differences in the amplitude of the long-duration tail come from the differences in the variation of the density along the line of sight, which are relatively small for the models we are considering. Considering the integrand in Eq.~(\ref{dflongduration}), the $x^2(1-x)^2$ term (from the radius of the microlensing tube), is largest at $x=0.5$, while the normalised density profile, $\tilde{\rho}(x)$, decreases monotonically with increasing $x$. For the models we are considering the integrand, and hence the sensitivity to the density profile, is largest for $x \approx 0.20-0.25$. 
  
The density profile for the PL halos, Eq.~(\ref{rhoPL}), is written in terms of a normalizing velocity, $v_{\rm a}$ rather than $\rho_{0}$ and the resulting expressions for ${\rm d} \Gamma/ {\rm d} \hat{t}$ are explicitly independent of $\rho_{0}$. The higher normalisation of the long-duration tails of the PL halos relative to the SH, arises from the potential of the disc not being taken into account when calculating the CO velocity distribution for the PL halos. As mentioned in Sec.~\ref{subsec:eventrate}, taking into account the disc contribution would increase the velocity dispersions in the Solar neighbourhood. If the SH were treated in the same way as the PL halos, then the asymptotic value of the circular speed would be $177 \, {\rm km \, s}^{-1}$ rather than $220 \, {\rm km \, s}^{-1}$. Figure \ref{fig:rate} shows the differential event rate for a halo composed entirely of CO with mass $M= 1 M_{\odot}$ for the SH, PLs B and C, and a SH with $177 \, {\rm km \, s}^{-1}$. The amplitudes of the long-duration tails of  PL B and C and the SH with $177 \, {\rm km \, s}^{-1}$ differ by less than ten per cent, showing that the rate of long-duration events depends only weakly on the DM density distribution.

\begin{figure}
    \centering
    \includegraphics[width=0.8\textwidth]{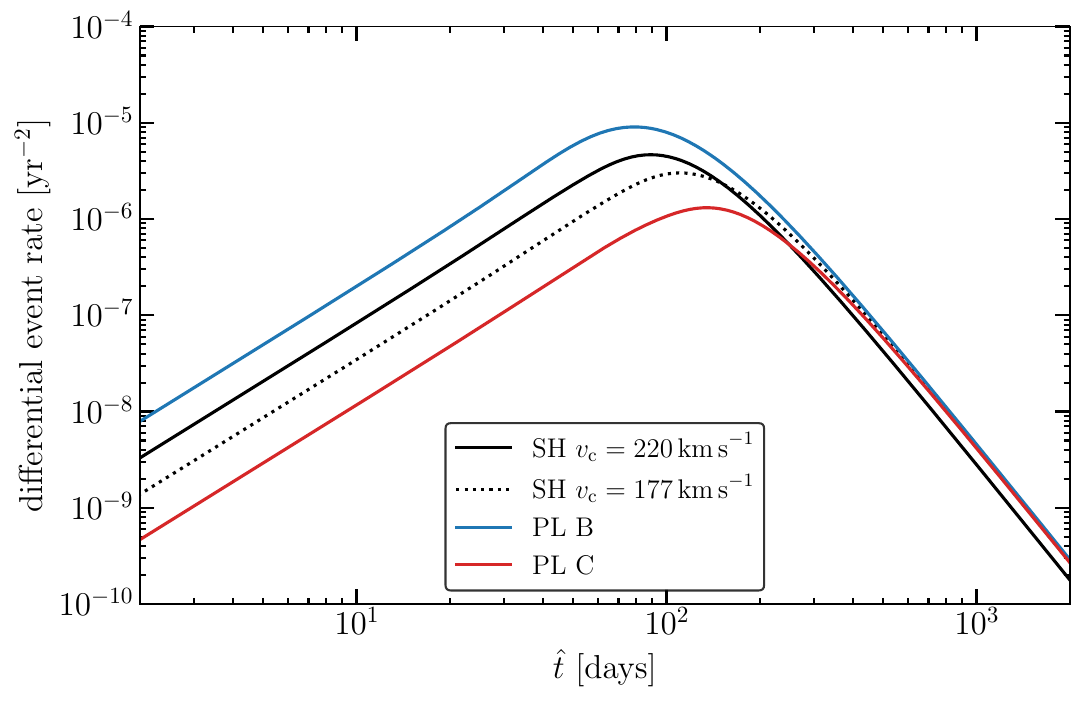}
    \caption{The differential microlensing event rate towards the LMC, ${\rm d} \Gamma/ {\rm d} \hat{t}$, as a function of Einstein diameter crossing time, $\hat{t}$, if the Milky Way halo is composed entirely of compact objects with mass $M= 1 M_{\odot}$ for the standard halo model in black and power law halos B and C in blue and red respectively.
The dotted black line is for the standard halo with $v_{\rm c}= 177 \, {\rm km \, s}^{-1}$ (rather than $220 \, {\rm km \, s}^{-1}$). 
    } 
    \label{fig:rate}
    \end{figure}

\subsection{Maxwellian velocity distribution}
\label{sec:gaussian}

\begin{figure}
    \centering
    \includegraphics[width=0.8\textwidth]{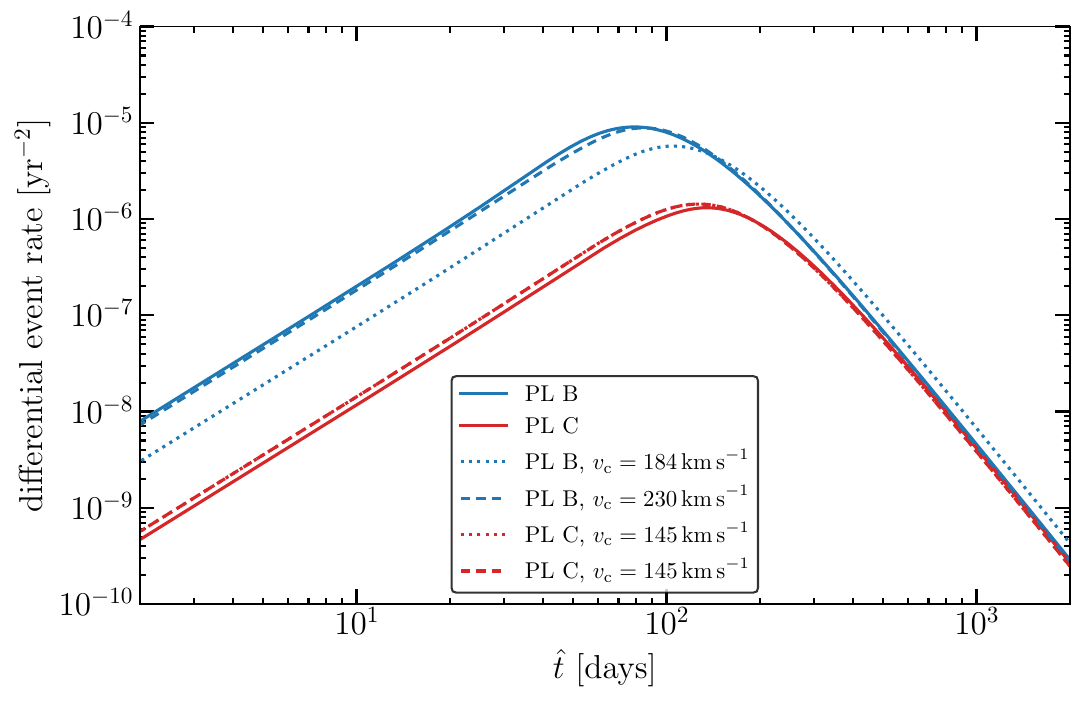}
    \caption{The differential microlensing event rate towards the LMC, ${\rm d} \Gamma/ {\rm d} \hat{t}$, as a function of Einstein diameter crossing time, $\hat{t}$, if the Milky Way halo is composed entirely of compact objects with mass $M= 1 M_{\odot}$ for power law halos B and C. The solid lines use the full expression for the differential event rate from Appendix B of Ref.~\cite{MACHO:1994ovs} (as in Fig.~\ref{fig:rate}), while the dotted and dashed lines use Eq.~(\ref{df}), which assumes a Maxwellian velocity distribution, and the density profile of the PL halos, Eq.~(\ref{rhoPL}). The dotted lines use the values of $v_{\rm c}$ at the Earth's position ($x=0$, $r=8.5 \, {\rm kpc}$), $v_{\rm c} = 184$ and $145\, {\rm km \, s}^{-1}$ for PL B and C respectively (see Fig.~\ref{fig:vc}). The dashed lines use values of $v_{\rm c}$ that roughly match the typical value for the range of radii probed by LMC microlensing,  $v_{\rm c} = 230$ and $145\, {\rm km \, s}^{-1}$ for PL B and C respectively. Note that for PL C the local ($x=0$) and typical values of $v_{\rm c}$ happen to be the same. } 
    \label{fig:fmaxwellian}
    \end{figure}

In Fig.~\ref{fig:fmaxwellian} we compare the differential event rates (for $M= 1 M_{\odot}$) for the PL halos with density profile given by Eq.~(\ref{rhoPL}), calculated using the standard expression, Eq.~(\ref{df}), which assumes a Maxwellian $f({\bf v})$, with those previously found using the accurate expressions in Appendix B of Ref.~\cite{MACHO:1994ovs}. We use values of $v_{\rm c}$ for the Maxwellian velocity distribution based on the findings in Sec.~\ref{sec:tails}. Short-duration events are mostly due to COs close to the source, where $v_{\rm c} = 184$ and $145\, {\rm km \, s}^{-1}$ for PL B and C respectively. Long-duration events are caused by COs moving along the line of sight, and for the range of radii probed by LMC microlensing the typical value of $v_{\rm c}$ is (roughly) $230$ and $145\, {\rm km \, s}^{-1}$  for PL B and  C respectively. The rotation curve of PL C peaks at a radius slightly largely than the Solar radius before declining, and the local and typical values of $v_{\rm c}$ are similar, while the rotation curve of PL B rises monotonically (see Fig.~\ref{fig:vc}). 

The long-duration tails of the differential event rates calculated assuming a Maxwellian $f({\bf v})$ with typical values of $v_{\rm c}$ agree with those from the full calculation fairly well (less than $10\%$ difference). This indicates that the detailed shape of the velocity distribution does not have a large effect on the rate of long-duration events (and hence exclusion limits for light COs). Surprisingly for the short-duration tail, for PL B (where the local and typical values of $v_{\rm c}$ are very different due to its rising rotation curve) the typical $v_{\rm c}$ value gives better agreement with the full calculation than the local $v_{\rm c}$, despite the majority of short-duration events being caused by local COs. However, this appears to be a coincidence; it is not the case for other values of $\beta$. We conclude that the amplitude of the short-duration tail of the differential event rate (and hence the exclusion limits on high-mass COs) depends significantly on the local velocity distribution of the COs.

\section{Summary}
\label{sec:summary}

Microlensing surveys of stars in the LMC place tight constraints on the fraction, $f$, of the MW halo in the form of planetary or Solar mass COs, including PBHs~\cite{Mroz:2024wag,Mroz:2024mse,Mroz:2024wia}. COs with mass in the range $10^{-9} \lesssim M/M_{\odot} \lesssim 10^{4}$ can not make up all of the DM, while those with mass $10^{-8} \lesssim M/M_{\odot} \lesssim 10$ are limited to $f \lesssim 0.01$.

The microlensing differential event rate (i.e.~the rate at which events of a given timescale occur), depends on the CO mass, density and velocity distributions. Uncertainties in the CO density and velocity distributions therefore lead to uncertainties in the constraints on the abundance of CO. We have focussed on understanding the dependence of the differential event rate, and hence the constraints, on the 
CO density and velocity distributions. This is a necessary step towards an accurate calculation of the constraints. We assume that the DM is smoothly distributed and use Evan's PL halo models~\cite{1994MNRAS.267..333E} as benchmarks, as they have tractable analytic expressions for the differential event rate which do not assume that the CO velocity distribution is Maxwellian~\cite{MACHO:1994ovs}.

The differential event rate has power-law tails, $\propto \hat{t}^{2}$ and $\hat{t}^{-4}$ for short and long-durations respectively~\cite{Mao:1996hr}. The typical event duration is proportional to $M^{1/2}$, so the constraints on low- and high-mass COs depend on the long- and short-duration tails respectively. Long-duration events are due to COs moving close to the line of sight~\cite{Mao:1996hr}. Their rate depends mainly on the integral of the CO density within the microlensing `tube'. For the models we have considered, the rate of long-duration events, and hence the exclusion limits on light-COs, only vary by of order $10\%$.

Short-duration events are due to COs that are close to either the observer or source, where the radius of the microlensing tube is small~\cite{Mao:1996hr}. Since the DM density is significantly smaller at the LMC than at the Solar radius, CO objects close to the Solar radius dominate. We have found an analytic expression, for the case where the COs have a Maxwellian velocity distribution, which is accurate at the $1\%$ level. The rate of short-duration events, and hence the exclusion limits on moderate- and high-mass COs, depends strongly on the velocities of the COs; the faster the COs are moving, the higher the rate at which they enter the microlensing tube and the shorter the events they cause. For a Maxwellian velocity distribution, the event rate is proportional to $v_{\rm c}^{4}$.  
We also found that assuming a Maxwellian velocity distribution for the PL halo models can lead to an inaccurate calculation of the differential event rate for short-timescale events.

We conclude that an accurate calculation of the DM velocity distribution in the Solar neighbourhood, where the potential of the MW disc is non-negligible, is required to place accurate stellar microlensing constraints on the abundance of COs with mass $M \gtrsim 10^{-2} M_{\odot}$.


\section*{Acknowledgements}

AMG is supported by STFC grant ST/P000703/1. For the purpose of open access, the author has applied a CC BY public copyright licence to any Author Accepted Manuscript version arising. \\

{\bf Data Availability Statement} This work is entirely theoretical and has no associated data.

\bibliographystyle{JHEP}
\bibliography{PBH.bib}

\end{document}